\documentclass[%
 reprint,
superscriptaddress,
 amsmath,amssymb,
 aps,
]{revtex4-2}

\usepackage{graphicx}% Include figure files
\usepackage{dcolumn}% Align table columns on decimal point
\usepackage{bm}% bold math
\usepackage[export]{adjustbox}
\usepackage{xcolor}
\usepackage{ulem}
\begin{document}

\preprint{APS/123-QED}
\author{A. Cavaillès}
\email{adrien@lighton.io}
 \affiliation{%
 LightOn, 3-5 Impasse Reille, 75014 Paris, France.
}%
\author{P. Boucher}%
\affiliation{%
Quantum Metrology and Nano Technologies Division, INRiM, Strada delle Cacce 91, 10153 Torino, Italy. 
}%
\author{L. Daudet}%
\affiliation{%
 LightOn, 3-5 Impasse Reille, 75014 Paris, France.
}%
\author{I. Carron}%
\affiliation{%
 LightOn, 3-5 Impasse Reille, 75014 Paris, France.
}%
\author{S. Gigan}%
\affiliation{%
 LightOn, 3-5 Impasse Reille, 75014 Paris, France.
}%
%\email{sylvain.gigan@lkb.ens.fr}
\affiliation{%
Laboratoire Kastler Brossel, ENS–Université PSL, CNRS, Sorbonne Université, College de France, 24 Rue Lhomond, F-75005, Paris, France.
}%
\author{K. M\"uller}%
\affiliation{%
 LightOn, 3-5 Impasse Reille, 75014 Paris, France.
}%

\begin{abstract}
Reconfigurable linear optical networks are a key component for the development of optical quantum information processing platforms in the NISQ era and beyond.
We report the implementation of such a device based on an innovative design that uses the mode mixing of a multimode fiber in combination with the programmable wavefront shaping of a SLM. The capabilities of the platform are explored in the classical regime. For up to 8~inputs and a record number of 38~outputs, we achieve fidelities in excess of $93\%$, and losses below $6.5\textrm{dB}$. The device was built inside a standard server rack to allow for real world use and shows consistent performance for 2x8 circuits over a period of 10 days without re-calibration.
\end{abstract}
\maketitle
%%%%%%%%%%%%%%%%%%%%%%%%%%  body  %%%%%%%%%%%%%%%%%%%%%%%%%%
\section{Linear optical quantum computing}

Photons are excellent carriers of quantum information: They are resistant to decoherence and can easily propagate over large distances. Mature technologies for controlling their creation, propagation, and detection are now available, or on the cusp of becoming useful. They also allow to interface between other quantum platforms like ions~\cite{monroe_2014}, neutral atoms~\cite{pivovarov_quantum_2020}, spin~\cite{chen_polarization_2021} and superconducting qubits~\cite{das_interfacing_2017}. Taken together, these properties give them a unique standing in the development of applications in the NISQ era and beyond.

A recent implementation of optical quantum information processing (QIP~\cite{obrien_optical_2007}) demonstrated supremacy over classical computers for the first time, but only for one specific and fixed task~\cite{zhong_quantum_2020}.
%It has recently been demonstrated that NISQ optical quantum information processing (QIP) systems~\cite{obrien_optical_2007} can already offer significant speedup over classical computers for certain tasks~\cite{zhong_quantum_2020}.
One of the most promising avenues for practical and scalable optical QIP are programmable linear optical networks~\cite{kok_linear_2007, harris_linear_2018, bromley_applications_2020}.
They are needed to implement QIP both in circuit model~\cite{knill_scheme_2001} and cluster state approaches~\cite{raussendorf_one-way_2001, raussendorf_measurement-based_2003}, and could find applications in diverse fields such as quantum transport~\cite{peruzzo_quantum_2010, crespi_anderson_2013, sparrow_simulating_2018, qiang_implementing_nodate}, quantum repeaters~\cite{lee_quantum_2020}, and quantum machine learning~\cite{shen_deep_2017, steinbrecher_quantum_2019, feldmann_parallel_2021, saggio_experimental_2021}.
Photonic integrated circuits are a popular choice for the implementation of such systems~\cite{matthews_manipulation_2009, carolan_universal_2015, flamini_photonic_2018, taballione_universal_2021, taballione_20-mode_2022}. Here the linear circuit is implemented via the propagation of light through a series of beam splitters and phase shifters in an integrated waveguide architecture. These systems see steady improvement, but their scalability is currently hindered by fabrication complexity, thermal stability and electrical control. 

3D wave mixing in complex media in conjunction with wavefront shaping is a promising alternative approach. The implementation of a simple beam splitter~\cite{huisman_programming_2014, huisman_programmable_2015, wolterink_programmable_2016, defienne_two-photon_nodate} as well as more complex linear circuits~\cite{leedumrongwatthanakun_programmable_2020, goel_inverse-design_2022} have been demonstrated. 
As any n-input, m-output circuit can be implemented in principle as long as the number of controlled modes is greater than $n*m$~\cite{leedumrongwatthanakun_programmable_2020_SI}, and because waveguides with propagation modes numbering well beyond the tens of thousands are readily available, the platform shows great potential for scalability.
%and with the availability of waveguides exhibiting more than 100,000 propagation modes, the platform shows great potential for scalability.}
%\textcolor{blue}{As any n-input, m-output circuit can be implemented in principle as long as the number of controlled modes is greater than $n*m$~\cite{leedumrongwatthanakun_programmable_2020_SI}, this type of system promises scalability beyond what can reasonably be achieved with integrated circuits.}
Furthermore, there is the possibility to interface with a wider range of optical systems of various wavelengths including outside the NIR/telecom range. Demonstrations so far have been proofs of concepts in a laboratory environment. To enable their use in the ever-expanding context of NISQ, QIP necessitates more integrated and transportable designs. Furthermore, while this approach has been tested for circuits with up to 2 input ports, many applications require a larger circuit, in particular when working with dual-rail encoding~\cite{bartolucci_creation_2021}.

In this paper, we report on the realization of a multimode-fiber-based optical linear processor installed in a standard server rack. After a classical characterization it is shown to allow for arbitrary linear quantum circuits with size up to 8x38, fidelities above $93\%$ and losses below $6.5~ \textrm{dB}$. We describe the so-called QORE processor and its working principle before providing a study of its performance in the classical regime in terms of circuit fidelity, stability and coupling.

\section{Qore: principle}

\begin{figure}[h!]
\includegraphics[width=0.5\textwidth]{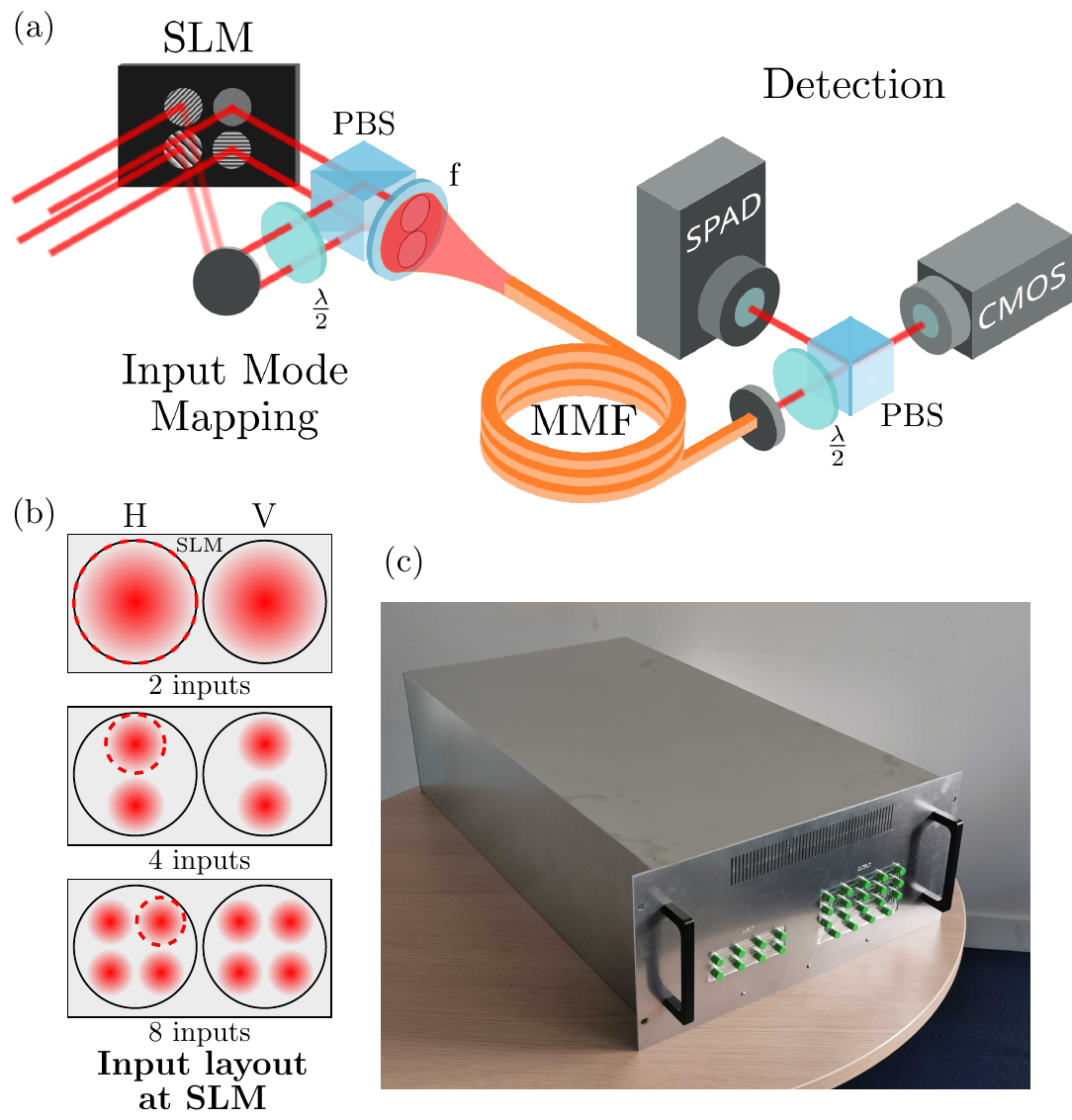}
\caption{\label{fig1}\label{fig1}\textbf{Device and principle}. (a): \textit{Operating principle}. Incoming input modes are mapped controllably onto the spatial modes of a multimode fiber using a spatial light modulator (SLM).
%Both polarizations are used as input through mixing using a polarizing beam splitter (PBS).
Orthogonal polarizations can be combined via a polarizing beam splitter (PBS) to double the number of inputs.
A calibration phase where the output field is monitored while applying various phase masks on the SLM allows the computation of the MMF's transfer matrix TM. Its inverse is then used to implement controlled circuits characterized both in the classical regime via either a CMOS camera or single-photon detector array (SPAD). (b): \textit{Input scaling}. 
Disposition of beams at the SLM level depending on the input configuration. The SLM is in the Fourier plane of the MMF and all incoming light has to fit in the black circle which corresponds to the fiber numerical aperture. Increasing the number of inputs lowers the number of accessible modes in the fiber as input beams have to be shrunk. Dashed red circle: size and location of the single input beam implemented to test a given input configuration.
(c): \textit{Prototype device}. Mechanical and thermal stability have been optimized for a rack-ready system.}
\end{figure}

The operating principle of the device is shown in Fig.~\ref{fig1}(a). The process harnesses the polarization and spatial mode-mixing properties of a multimode fiber (MMF) in conjunction with the wavefront shaping capabilities of a spatial light modulator (SLM) to implement the desired optical circuit $\mathcal{L}$. The wavefronts of incoming input fields on the left of the figure are shaped through the application of suitable phase masks on the SLM before being coupled into the MMF. The output fields are then verified to follow $\mathcal{L}$, both in the classical regime or using single photon detectors.

Each incoming input field is mapped onto the appropriate modes of the MMF in two steps. First the measurement of the MMF's transfer matrix (TM) is performed following a phase-stepping holographic method. 
In this method, the fiber input mode basis is probed by displaying on the SLM gratings of various directions and periods added to a global dephasing, and the resulting output fields are recorded with the chosen detection system (see \cite{popoff_measuring_2010, cizmar_shaping_2011} for more details).
%followed by a fine-tuning via machine learning optimization.
The TM can then be inverted --taking advantage of its unitarity, we compute in practice its conjugate transpose-- and applied to the target output coupling to obtain the appropriate phase mask to be displayed on the SLM.
In a final step, we fine-tune the SLM mask using a differential evolution algorithm to achieve the best performance. Repeating the procedure for each input leads to the realization of a circuit $\mathcal{L}_{\textrm{exp}}$ that can be compared with the targeted $\mathcal{L}$. The possibility of implementing a wide range of circuits is due to the largely isotropic polarization and spatial mode mixing taking place in the MMF, the many availble degrees of freedom on the SLM as well as the unitary nature of the TM.

The concept has been presented and used in \cite{leedumrongwatthanakun_programmable_2020, goel_inverse-design_2022} in the context of up to two optical inputs distinct in polarization. Here we report the extension of the system to  8 inputs as well as its characterization.
Each input is incident on the SLM (Holoeye Pluto-2-NIR-080) as a gaussian beam before the appropriate phase masks are applied. The SLM is in the Fourier plane relative to the input facet of the MMF. Figure~\ref{fig1}(b) shows the input beams in that plane in red. The MMF aperture (shown as a black circle) limits the acceptable range of input modes.
One can see that the available area for each input decreases as their number grows. Consequently the number of MMF modes per input decreases, leading to a change in performance that we characterize in section~\ref{sec:cap}. 
% This increase leads to a diminution of the number of spatial modes of the MMF that can be exploited for each input. This can be understood by looking at Fig.~\ref{fig1}(b).
% Each input beam arrives at the SLM with a gaussian wavefront before being applied appropriate phases masks.
% The SLM is in the Fourier plane relative to the input facet of the MMF and in this figure the wavefront of inputs are shown in that plane in red.
% The MMF aperture (shown as a black circle) limits the range accepted of accepted incoming modes.
% One can see that the available area for each input decreases as their number rises, although with a diminishing effect going above 4 inputs.
As an illustration, Fig.~\ref{fig1}(a) shows the configuration corresponding to 2 input railings per polarization. We tested the system for configurations going up to 8 inputs.

% Along with the input basis the realisation of optical circuits $\mathcal{L}$ requires one to define the corresponding output basis or railings.
Along with the input basis, the realization of optical circuits requires the definition of an output basis, or railings.
Ultimately, these output railings aim to be matched to detection or propagation channels. In our case the quality of circuits is tested using two kinds of detection systems: a CMOS camera (Basler aca2000) and an array of 23 avalanche photodiodes (Pi Imaging SPAD23). Both are imaging the output facet of the MMF. The rotation of the half-waveplate at the output allows us to monitor both polarizations either with the camera or the single-photon detector array. With the camera, the output modes are defined as a set of zones on the sensor with an area on the order of the speckle grain size. 
%These are shown in Fig.~\ref{fig:fidcoup}~(a) as red circles. 
In the case of the SPAD23, a total of 46 output modes are defined as the active areas of the 23 detectors, monitored on both polarizations.
With these definitions set, one can see that we are able to evaluate the performance of the system in a variety of input-output number configurations $n$x$m$ using a single input beam: To adapt to the input configuration $n$, we simply suitably offset and resize the beam on the SLM plane following the dashed-red circles in Fig~\ref{fig1}(b). We can then monitor $m$ detection modes to match the output setting.  

\section{\label{sec:cap}Qore: capability}
\begin{figure*}
\center
\includegraphics[width=0.8\textwidth]{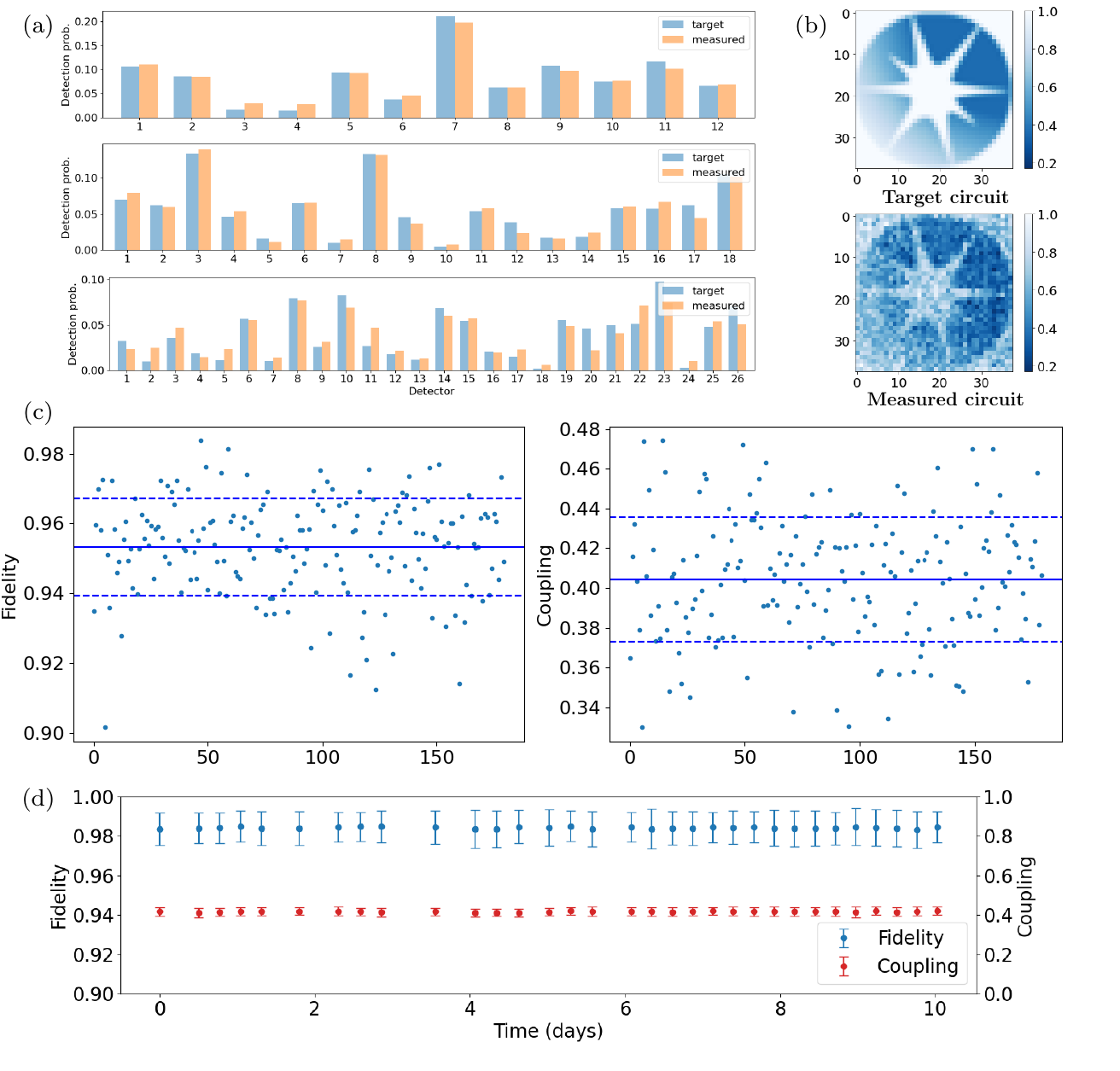}
\caption{\label{fig:fidcoup}\textbf{Performance of the device}: (a): Target vs measured detection probability for 3 example circuits using the SPAD23 detector array with 12, 18, and 28 outputs. (b): Example implementation of a 38x38 non-unitary target circuit. Each single input to 38 outputs amplitude distribution (corresponding to the lines of the matrix) has been measured separately in a 2-input configuration. The data-sets are re-normalized by their maxima. We measure a statistical fidelity of $98.6\%$.
(c) Fidelity and coupling for 180 randomly distributed unitary circuits in 2x26 input-output configuration. Full line: mean, dashed lines: mean $\pm1\sigma$. (d): Evolution of both parameters over 10 days without re-calibration. This characterization was done with the camera, in a 2x8 input-output configuration for 160 circuits. The error bars correspond to 1 standard deviation from the mean. Following a single-hour calibration the fidelity and coupling see no significant degradation over the period.}
\end{figure*}
In Figures~\ref{fig:fidcoup}(a) and \ref{fig:fidcoup}(b), we present experimentally implemented circuits as measured using the SPAD23 in the configuration corresponding to 2 inputs and variable output number settings. Fig~\ref{fig:fidcoup}(b) in particular illustrates the control we have on the circuit generation in the case of 38 outputs.  As shown on top, we used here as a target circuit a non-unitary matrix reproducing the LightOn logo. Each line of the matrix was implemented in sequence and then measured to obtain the data-set displayed on the bottom.
From this information, the quality of the implemented optical circuits can be assessed through two measures: circuit fidelity and transmission loss. The fidelity is an indication of how close the implemented circuit $\mathcal{L}_{exp}$ is to the targeted one $\mathcal{L}_{target}$ and is expressed as the trace of the product between the two: $\mathcal{F}=\frac{1}{D}\textrm{Tr}(\mathcal{L}_{exp}\mathcal{L}_{target}^{\dag})$, with D the dimension of $\mathcal{L}_{exp}$. The upper bound of $\mathcal{F}$ is one. In practice we record the related statistical fidelity defined as $\mathcal{F}=\frac{1}{D}\textrm{Tr}(\lvert \mathcal{L}_{exp}\rvert \lvert \mathcal{L}_{target}\rvert)$ in the output subspace, i.e. when correcting $\mathcal{L}_{exp}$ for global transmission losses through its normalization. In the case of Fig.~\ref{fig:fidcoup}(b), for example we find an average statistical fidelity of $98.6\%$.
%It can be conveniently expressed in terms of the amplitude probabilities of the target and measured circuits over output modes $j$ from input mode $i$, respectively $p^{i,j}_{target}$ and $p^{i,j}_{exp}$  as $\mathcal{F} = \sum\limits_{i,j} \sqrt{p_{exp}^{i,j}p_{target}^{i,j}}$. 

The second measure is the transmission loss of a circuit, defined as the proportion of intensity at the input that is coupled into the targeted output modes. The losses from the input up to the output of the fiber are directly measured using a Thorlabs PM100 power-meter. The SLM introduces losses through several properties. The $95\%$ reflectivity of the screen directly translates into $5\%$ of losses independently of the circuit. However, the losses resulting from its limited fill factor ($93\%$) and diffraction efficiency will change with the displayed phase mask. The power-meter is therefore used to monitor the transmission through the fiber on a circuit-by-circuit basis. The remaining coupling into the output railings is assessed through an initial calibration procedure relating the power-meter and SPAD23 system detection efficiencies (SDE). The ratio between the two SDEs is related to the total number of photons reaching the detector plane for a given measured power on the PM100. As we only have information on the active areas of the SPAD23, this value is estimated from a fitting procedure on the count rates observed on the detectors. From that fit we obtain the target number of counts that would be measured for a hypothetical detector with $100\%$ fill-factor. We find that the procedure leads to consistent estimations of the ratio between SDEs for various beam waists and angles with $6\%$ standard deviation overall.
%differently depending on the detection system used. 
%In the case of the CMOS camera it is obtained as the ratio between the measured integrated intensity in the output modes and the overall integrated intensity. 
%For the SPAD23 we first perform a calibration procedure relating the power-meter and SPAD23 system detection efficiencies. 
%In practice, this calibration is estimated to be accurate up to $6\%$ of standard deviation. 
The last source of loss is due to reflections at every interface and is measured to be $10\%$. Since the application of anti-reflection coatings would greatly reduce this loss, we provide in table~\ref{tab:results} values both with and without correction from reflection losses.
Following this methodology, we implemented between $100$ and $120$ unitary circuits ${\mathcal{L}_k}$, randomly chosen from the Haar measure, per input-output setting. For each circuit the statistical fidelity in the output subspace and transmission loss was measured. In practice, the final fine-tuning of parameters was performed such as to maximize the statistical fidelity without correction for transmission losses. This represents the appropriate compromise between circuit fidelity and transmission losses for most applications. However this last optimization can be tuned so as to favour one of the parameters if necessary.
%From this methodology we implemented a number (700 circuits per input-output setting) of randomly distributed (along the Haar measure) unitary circuits ${\mathcal{L}_k}$ on the system and measured the corresponding fidelity and transmission loss.
The light source used is a superluminescent diode centered on $810~\textrm{nm}$, and filtered by a $2~\textrm{nm}$ FWHM spectral filter. As we use a single input for the characterisation and unitary circuits are square, we implemented in sequence each line of $\mathcal{L}_k$ and then averaged the values to obtain the overall circuit fidelity and loss. These two measures are shown in figure~\ref{fig:fidcoup}(c), in the case of 2 inputs and 26 outputs when measured using the SPAD23. We find an average fidelity of $95.3\%\pm1\%$ and total coupling from input to detection mode averaging at $40\%$ (which corresponds to $4~\textrm{dB}$ losses), and $45\%$ when correcting for the reflection losses. The process has been repeated for a number of configurations and results are summarized in Table~\ref{tab:results}. We obtain high values of statistical fidelities averaging above $93\%$ for all considered input-output configurations and observe low variation in quality depending on the circuit at a fixed input size. This compares favorably with many alternative methods based on integrated circuits~\cite{carolan_universal_2015, taballione_universal_2021} at comparable output sizes and is close to recent demonstrations~\cite{taballione_20-mode_2022} for $26$ outputs. The ease of scaling to a larger number of outputs is one significant advantage of the 3D wave-mixing approach, as leading re-configurable integrated systems are currently limited to our knowledge to 20x20 circuits~\cite{taballione_20-mode_2022}.

While the measured transmission loss is 
%competitive as well with integrated systems in which coupling into the waveguide is a significant source of loss, it is 
still high for specific usages such as quantum advantage demonstrations ($\sim 2~\textrm{dB}$ in \cite{zhong_quantum_2020}), it can be mitigated through various avenues. First, through the addition of anti-reflection coatings on the fiber; second, the independent control of two polarization channels at the input --- instead of one in the current system --- would lead to a decrease of losses. Another possibility would be the addition of input beam shaping to improve the packing at the SLM level, so that the number of spatial modes used by each input railing is greater. One should note as well that these results are intrinsically linked to the specific properties of the detector. We used a commercial system that is therefore not specifically designed for the procedure. In particular, increasing the fill-factor of the detection device beyond the SPAD23's $23.5\%$ would lead to an improvement in circuit quality. The quality of implementation of the circuit will also depend on the temporal modes of the quantum states used. As all measurements have been performed at $810\pm 2~\textrm{nm}$, the quality of circuits implemented for narrower circuits can only improve from our results. 
On the other hand, if a degradation in quality can be expected for broader sources, we note however that it can be mitigated through the adaptation of the multimode fiber properties, in particular its length. 
As it only monitors the circuits from a single input, our characterization method cannot take into account the background noise introduced by eventual other inputs coupled into the system. This effect could however be mitigated through an additional fine-tuning once all input masks have been obtained separately.  

\begin{table*}
%\begin{ruledtabular}
\begin{tabular}{lcccc}
 &  &\multicolumn{1}{c}{\textbf{2 inputs}}&\multicolumn{1}{c}{\textbf{4 inputs}}&\multicolumn{1}{c}{\textbf{8 inputs}}\\ \hline
 \rule{0pt}{3ex}   
\textbf{Fidelity} & \textbf{14 outputs} & $97.1\%\pm 1\% $ & $96.6\%\pm 1\%$ & $96.3\%\pm 1\%$ \\
& \textbf{26 outputs} & $95.3\%\pm 1\%$ & $94.7\%\pm 1\%$ & $94.2\%\pm 1\%$  \\
& \textbf{38 outputs} & $93.7\%\pm 1\%$ & $93.5\%\pm 1\%$ & $92.9\%\pm 1\%$ \\\hline
\rule{0pt}{3ex}   
\textbf{Losses} & \textbf{14-38 outputs} & $4.0 (3.5^*)\textrm{dB}\pm 0.5\textrm{dB}$ & $5.1 (4.6^*)\textrm{dB}\pm 1\textrm{dB}$ & $6.2 (5.7^*)\textrm{dB}\pm 1.2\textrm{dB}$ \\
\end{tabular}
%\end{ruledtabular}
\caption{Measured statistical fidelity and losses for different input-output configurations as measured on the SPAD23 for more than 100 unitary circuits distributed randomly accross the Haar measure. *: Values estimated with the addition of anti-reflection coatings.}
\label{tab:results}
\end{table*}

Another important aspect of the system is its stability over time. In Fig.~\ref{fig:fidcoup}(d) we plotted the measured fidelity and coupling for a fixed set of 160 2x8 circuits over more than 10 days after a single initial calibration. As one can see, there are variations in quality but no significant degradation over the whole period. These measurements were obtained with the rack set on an optical table with no active vibration system, laminar flow or temperature control in place other than the lab room air conditioning. This level of stability can be therefore expected or exceeded in most settings where quantum sources or detection systems are used. The characterization of the transfer matrix takes about two hours per input mode to perform, meaning that the day-to-day use remains practical even for 8 inputs and beyond.

Two additional properties of the approach distinguish it from integrated alternatives. First, it is largely wavelength agnostic as the system can work over a wide range of wavelengths without specific adaptations, apart from the change of (anti-)reflection coatings and SLM model.
Second, the circuit reconfiguration speed is fixed by the refresh rate of the modulator no matter the circuit size. While integrated systems of comparable dimensions typically require reconfiguration time in the order of 1s, the rate is in our case close to $10~\textrm{Hz}$. In addition, ongoing developments of the SLM technology hold the promise of significantly faster updating. 
Commercial models exhibiting $700~\textrm{Hz}$ of refresh rate are already available while new architectures going to the MHz level have been proposed~\cite{peng_design_2019}. These adaptations would significantly broaden the capabilities of the platform, allowing for example the consecutive application of gates conditioned on the results of preceding ones.

\section{Conclusion}
In this paper, we have introduced and characterized a compact and linear optical processor based on complex mixing and wavefront shaping. We find that the platform brings significant benefits: advantageous scaling --- as exemplified by the realisation of a record 38-output circuit ---, high fidelities and competitive reconfiguration rate, while maintaining week-long stability for a 2-inputs, 8-outputs configuration. It represents as such a novel tool for optical QIP that could be used as a convincing alternative or in conjunction with integrated systems. The asymmetry of the implemented circuits suggests one exciting prospect in particular: the generation of high-dimensional entanglement from few single photon states, thus allowing the interconnection between many quantum platforms.

\section{Acknowledgments}
We acknowledge support from EU Horizon 2020 FET-OPEN OPTOLogic (Grant No. 899794).

%%%%%%%%%%%%%%%%%%%%%%% References %%%%%%%%%%%%%%%%%%%%%%%%%

%%%%%%%%%% If using BibTeX:

%%%%%%%%%% If preparing manually:
% \begin{thebibliography}{1}
% \newcommand{\enquote}[1]{``#1''}

% \bibitem{Zhang:14}
% Y.~Zhang, S.~Qiao, L.~Sun, Q.~W. Shi, W.~Huang, L.~Li, and Z.~Yang,
%   \enquote{Photoinduced active terahertz metamaterials with nanostructured
%   vanadium dioxide film deposited by sol-gel method,}
%   {\protect\JournalTitle{Optics Express}} \textbf{22}, 11070--11078 (2014).

% \bibitem{OSA}
% {Optical Society}, \enquote{{OSA Publishing},}
%   \url{http://www.osapublishing.org}.

% \bibitem{FORSTER2007}
% P.~Forster, V.~Ramaswamy, P.~Artaxo, T.~Bernsten, R.~Betts, D.~Fahey,
%   J.~Haywood, J.~Lean, D.~Lowe, G.~Myhre, J.~Nganga, R.~Prinn, G.~Raga,
%   M.~Schulz, and R.~V. Dorland, \enquote{Changes in atmospheric consituents and
%   in radiative forcing,} in \enquote{Climate Change 2007: The Physical Science
%   Basis. Contribution of Working Group 1 to the Fourth assesment report of
%   Intergovernmental Panel on Climate Change,}  S.~Solomon, D.~Qin, M.~Manning,
%   Z.~Chen, M.~Marquis, K.~B. Averyt, M.~Tignor, and H.~L. Miler, eds.
%   (Cambridge University Press, 2007).

% \end{thebibliography}

\end{document}